\documentclass[floatfix,a4paper]{article}
\usepackage{amsmath,amssymb,graphicx}

\date{}

\begin{document}

\title{From quantum mechanics to the physical metallurgy of steels}
\author{Anthony T. Paxton}
\maketitle

\begin{center}
{\large\it
Department of Physics,\\ King's College London,\\ Strand, London WC2R 2LS, UK
}
\end{center}

\begin{abstract}
  In the last decade there has been a breakthrough in the construction
  of theories leading to models for the simulation of atomic scale
  processes in steel. In this paper the theory is described and
  developed and used to demonstrate calculations of the diffusivity
  and trapping of hydrogen in iron and the structures of carbon
  vacancy complexes in steel.
\end{abstract}

\def\br{{\bf r}}
\def\d{{\rm d}}
\def\aFe{$\alpha$-Fe}
\def\phip{{\tilde\phi_p}}

\section{Historical Introduction}
\label{sec_Intro}

In 1947 Geoffrey Raynor wrote ``\dots if metallurgists of this country
do not learn about the electron theories they may discover too late,
that the industry in other countries has benefited from knowledge of
modern theories and is developing new alloys.'' Hume-Rothery and
Raynor had understood perfectly from the work of Mott and Jones before
the war that to unravel the complexities of metal phase diagrams
required the powerful methods of {\it metal physics} that had arisen
out of the new quantum mechanics of Schr\"odinger and Heisenberg of
the 1920's~\cite{MottJones}.  But some 15 years later, Hume-Rothery
writes ``The work of the last ten years has emphasised the extreme
difficulty of producing any really quantitative theory of the
electronic structure of metals, except for those of the alkali
group.'' The breakthrough came just a few years after that with the
appearance of the {\it density functional theory} (DFT) of Kohn and
co-workers~\cite{HohenbergKohn,KohnSham} and for which Walter Kohn
shared the Nobel Prize in {\it Chemistry} (with John Pople) in 1998.

It was understood then, and is understood now, that what holds
together the atoms in a metal are the valence electrons. These act as
the ``glue'' which is ultimately responsible for the properties that
we exploit in metallurgy to produce alloys of prescribed hardness,
ductility, fracture resistance, and so on. What Hume-Rothery meant by
{\it electronic structure} is the stationary state solutions of the
Schr\"odinger equation for the electrons in condensed matter moving in
the (so called ``external'') potential of the fixed nuclei. At issue
in the 1960's was the awareness that this was fundamentally a {\it
  many electron} problem, requiring the knowledge of a wavefunction
$\Psi(\br_1,s_1,\br_2,s_2\dots\br_N,s_N)$ which depends on the
positions and spins of all the $N$ electrons involved. On the other
hand, every student learns quantum mechanics in terms of a
Schr\"odinger equation whose solution, $\psi(\br,s)$, is the
wavefunction of a single particle moving in an electrostatic
potential, $V(\br)$, which at most includes the potentials of all
other interacting particles in some average way. Because the
electron--electron interaction is due to the long ranged Coulomb force
it was thought inconceivable that a ``one-electron'' theory would
yield useful results for the electron gas in metals, and therefore it
was a bold move by physicists even to attempt this approach.

\subsection{Density functional theory}

The principal discovery in the DFT was that the ground state energy of
a many electron system is uniquely determined by the electron density
alone, meaning that it is not necessary to find the complicated many
electron wavefunction~\cite{HohenbergKohn}. Moreover it was found that
the many-electron Schr\"odinger equation could be cast exactly into
one-electron form if the potential $V(\br)$ is regarded as an {\it
  effective potential\/} including a contribution called ``exchange
and correlation'' which includes all the complicated
electron--electron interactions~\cite{KohnSham}. The problem is only
solvable, however, if one makes an approximation to this
exchange--correlation potential; this is the {\it local density
  approximation} (LDA) which asserts that every electron sees the
potential due to other electrons as if it were living in a uniform
electron gas, whose density is taken as that density which the
electron momentarily encounters.

For two reasons the DFT did not make immediate impact in the world of
metallurgy and solid state physics. Firstly the LDA was received with
great scepticism since it seemed to be sweeping a great deal of the
essential physics under the carpet. Secondly the solution of the
``self consistent problem'' requires a fast computer and complicated
computer program.\footnote{In casting the many-electron
problem into one-electron form it happens that the effective
potential comes to depend on the electron density which is the square
of the one-electron wavefunction. In that way the potential that
enters the Schr\"odinger equation depends on its solution. This means
that one needs to seek a {\it self consistent} solution to the problem
that can only be done with a computer.}
The first problems solved were to calculate the
electronic structure itself, especially energy bands and Fermi
surfaces; and it was not until it was possible to calculate the total
energy and compare crystal structures, calculate heats of formation
and elastic constants which could be compared to experiment that it
was realised that the LDA was an approximation that was much better
than it ought to have been. To jump some four decades ahead, it is now
fully accepted that the LDA is able to predict structural properties
of metals and alloys, to compute interatomic forces and to make
atomistic simulations, including molecular dynamics (MD) at least for
a few tens of atoms over a few picoseconds of simulation time.

\subsection{Tight binding theory}

The previous sentence gives the clue as to why the LDA cannot be
expected to be the final word in atomistic simulations. Evidently if
one wishes to make large scale MD simulations or molecular statics
relaxations to study processes such as diffusion, dislocation core
structures and mobilities, the current recourse is to so called {\it
classical\/} interatomic potentials, for example the well known
embedded atom method~\cite{Daw84}. Classical potentials display a number of
unrepairable difficulties. They cannot describe magnetism (essential
for the study of iron and steel); they cannot account for metals with
negative Cauchy pressure; and in all but a few exceptional cases they
fail to reproduce properly the core structure of dislocations in
\aFe~\cite{Mrovec09}.

There is however a middle way, and this is the tight binding (TB)
approximation. TB is a fully quantum mechanical approach to the
electronic structure calculation and yields total energy and
interatomic forces that can be used in MD simulations. On the other
hand it is much faster computationally than the LDA because the
hamiltonian is not constructed from first principles; instead a look
up table of hamiltonian matrix elements is created by fitting certain
fundamental parameters to experiment and to accurate density
functional calculations~\cite{Finnis03,Paxton09}. There is a deeper
fundamental theory behind the TB approximation that goes beyond the
notion of a fitted semi-empirical scheme. A number of authors have
shown that the TB approximation can be derived directly from DFT using
the notion of first order and second order expansions of the
Hohenberg--Kohn total energy functional in the difference between some
{\it assumed\/} input charge density and the actual self consistent
density~\cite{Sutton88,Foulkes89,Finnis03}. TB can describe itinierant
magnetism in transition metals using the self consistent Stoner-Slater
theory~\cite{Paxton08} and successful models have now been constructed
for \aFe\ and $\gamma$-Fe, both pure, and containing interstitital
impurities, hydrogen and carbon~\cite{Paxton10,Paxton13}.

\section{Atomistic modelling of processes in steels}

It is only in the last decade that metal physicists have come around
to tackling the problem of interstitials in Fe. And only once the case
of C in Fe is solved can one be said to be able to model processes in
steel. There are a few classical potentials, to describe both H and C
in \aFe~\cite{EAC09,Becquart07,Lau07}, but these are prone to the same
difficulties as those outlined above for pure iron, in addition to
which the number of fitted parameters rises dramatically so that for
example the H--Fe interactions require more than 40 parameters with
which to be described~\cite{EAC09}. Therefore it is very instructive
to determine what can be done using the ``middle way'' tight binding
approximation. This is followed through in the examples in the next
sections.

\subsection{Hydrogen in \aFe}

\begin{figure*}
  \caption{\label{figure1} The minimum energy path for proton
    migration in \aFe. The open circles show the energy as a function
    of proton position in \AA\ as it moves in the potential well
    around the tetrahedral interstice. The dotted line is a parabola
    fitted to the points. The curve labelled $\phip$ is a graph of the
    gaussian wavefuncion of the proton moving the the harmonic
    oscillator potential~\cite{Cassels}. A horizontal line marks the
    zero point energy relative to the bottom of the well. This is the
    energy of the ground state of the proton, which in the periodic
    potential of the lattice of tetrahedral sites forms a band of
    states of width on the order of the hopping integral $J$
    (equation~(\ref{eq_J})) as indicated by the horizontal dotted
    lines.}
\begin{center}
\includegraphics[scale=0.4, viewport=61 177 557 633, clip]{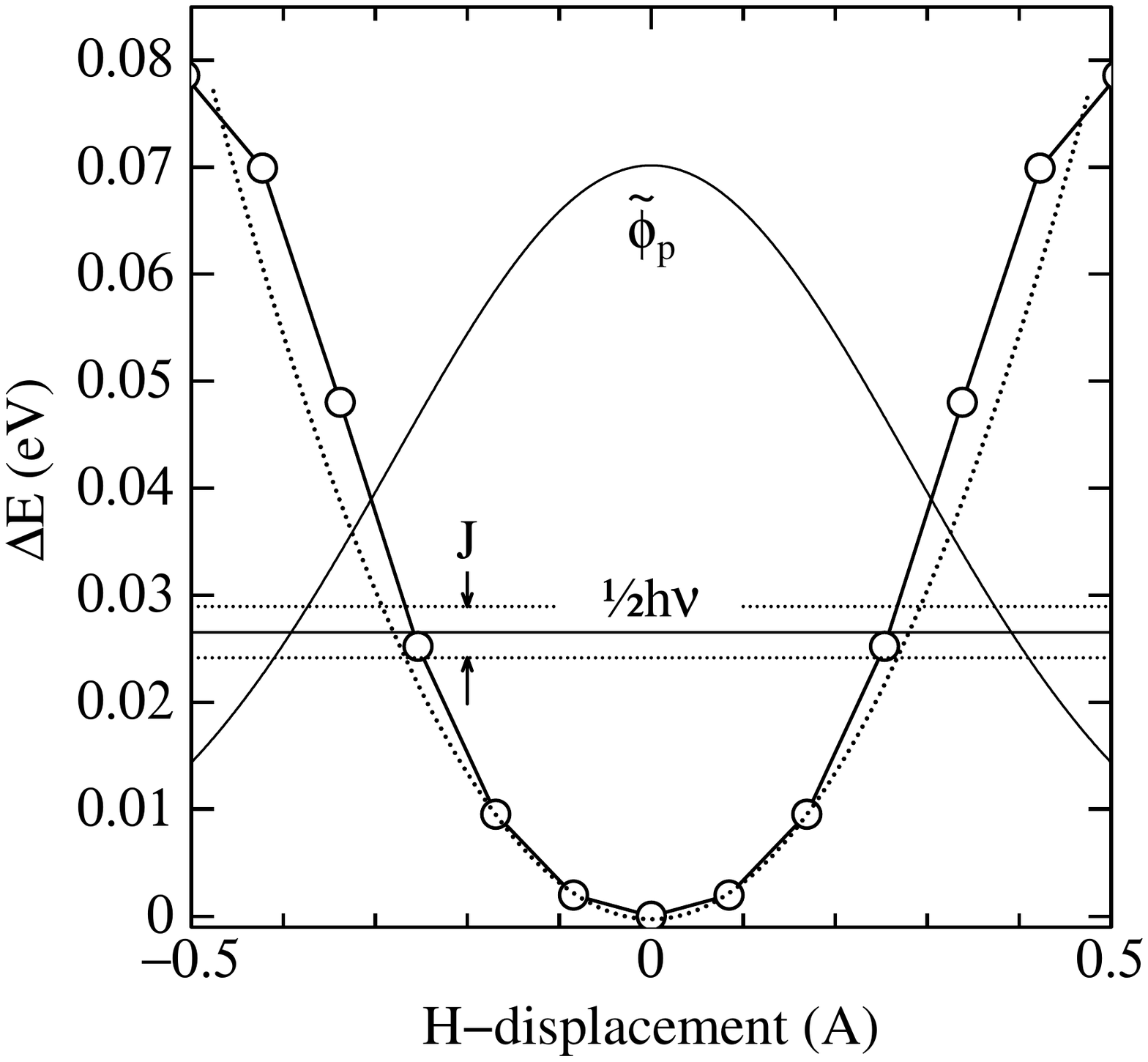}
\end{center}
\end{figure*}

In some sense H is the most interesting of the intersitial impurities
because of its small atomic mass. This fact coupled with the geometry
of the bcc lattice leading to a rather small activation energy for
diffusion between tetrahedral interstices means that the hydrogen
atom, or proton, cannot in general be treated as a classical
particle. A hydrogen molecule is characterised by a rather strong
covalent bond arising from the splitting of the atomic H-$1s$ level
into bonding and antibonding molecular orbitals; only the bonding
levels are occupied (by two electrons) and antibonding oribitals are
unoccupied~\cite{Coulson}. This situation changes when the H$_2$
molecule finds itself at the surface or within a transition metal. In
that case the H-$1s$ levels are lower in energy than the bottom of the
transition metal $d$-bands and so both the bonding and antibonding
levels become occupied---the covalent bond breaks apart and the H~atom
is free to move about the crystal lattice. It makes sense to think of
the H-atom as a freely moving proton that has given up its electron to
the energy bands of the solid. Since the H-$1s$ orbitals are fully
occupied it is tempting to think of the H-atom as {\it negatively}
charged, that is H$^{-}$, but this is not really helpful as any charge
in a metal is very efficiently screened by the conduction
electrons. The proton in solution in \aFe\ occupies the tetrahedral
interstices. The quantum nature of the proton is illustrated in
figure~\ref{figure1}. Here the TB method~\cite{Paxton10} has been used
to map out the so called minimum energy path~\cite{Henkelman00} for
the proton to move from its tetrahedral site to a neighbouring
site. In the process the proton passes through a maximum in the energy
along that path; in fact this is a saddle point in the total energy
landscape as understood from classical theory of
diffusion~\cite{Vineyard57}.

The dotted line in figure~\ref{figure1} is a parabola fitted to the
calculated energy showing that to a good approximation the proton is
moving in a quadratic potential. This means that it possesses the
wavefunction of a harmonic oscillator, namely the normalised
gaussian~\cite{Cassels} 
$$
\phip=\left({{2\alpha}\over{\pi}}\right)^{1/4}
      \,e^{-\alpha x^2}
$$
where
$$
\alpha ={{1}\over{2}}{{m\omega}\over{\hbar}}
$$
Here $x$ is the displacement from the equilibrium position, $m$ is
the mass of the proton and $\omega$ is the angular frequency of
vibration which can be extracted easily from the curvature of the
parabola. In fact the frequency of vibration is found here to be 
$$
\nu=\omega/2\pi=1.29\times 10^{13}\hskip 6pt \hbox{s$^{-1}$}
$$
There is a hamiltonian matrix element, or tight binding ``hopping
integral'', which connects the gaussian proton wavefunctions in
neighbouring interstices and this is~\cite{Flynn70,Flynn71e}
\begin{equation}
J = \frac{1}{4}\hbar\omega\left(3+\alpha R^2\right)
    \,e^{-{{1}\over{2}}\alpha R^2}
\label{eq_J}
\end{equation}
where $R=a/\sqrt{8}$ is the separation between neighbouring
tetrahedral interstices and $a$ is the bcc lattice constant. In this
case $\alpha R^2=0.17$.

In figure~\ref{figure1} is shown the proton wavefunction, $\phip$, the
zero point energy, $\hbar\omega/2$, and the magnitude of the hopping
integral $J$. Several points are notable. The zero point
energy is large on the scale of the energy barrier so that the proton
does not occupy the energy at the bottom of the potential well as
would a classical particle, instead even in the ground state the
proton possesses an energy nearly half way up to the saddle
point. If we think of the energy curve as a portion of a periodic
potential, then it is clear that at zero temperature, the proton
occupies a Bloch state or energy band whose width is on the order of
$J$. Both these facts lead to the conclusion that a {\it classical\/}
approach to the diffusivity of H~in \aFe\ is likely to lead one into
error. 

\subsubsection{Hydrogen diffusion in \aFe}

In the classical approach to diffusion, the so called transition state
theory, one is interested in the {\it classical} rate constant,
$\kappa^{\rm cl}$, which can be expressed in terms of a ratio of
partition functions~\cite{Vineyard57,ChristianI}. We are interested in
the ``reactant state'' in which the diffusing atom occupies a lattice,
or interstitial, site and its partition function is denoted $Z_R^{\rm
  cl}$. The second partition function of interest concerns the
particle at the saddle point. Following Vineyard~\cite{Vineyard57} we
construct a {\it constrained\/} partition function, $Z_c^{\rm
  cl}(q^*)$, in which the asterisk on the reaction coordinate, $q$,
conveys the fact that it is confined to the dividing surface in
configuration space: this is the surface perpendicular to the minimum
energy path. This reduces the dimensions of the configuration space
available to the particle by one. We then
have~\cite{Vineyard57,Voth93}
\begin{equation}
\kappa^{\rm cl} = \frac{1}{2} v\, \frac{Z_c^{\rm cl}(q^*)}{Z_R^{\rm cl}}
\label{rate_constant}
\end{equation}
and $v$ is interpreted as the particle's velocity as it approaches the
barrier. If this is taken from a Maxwell distribution at temperature $T$, then
$$
v = \sqrt{\frac{2}{\pi m\beta}}
$$
where $\beta=1/kT$ and $k$ is the Boltzmann constant.

\begin{figure*}
  \caption{\label{figure2} Diffusion coefficients of H in \aFe\
    calculated by path integral quantum transition state theory (PI
    QTST)~\cite{Katzarov13}.  The bands of data represent an
    assessment by Kiuchi and McLellan~\cite{km} of hydrogen gas
    equilibration experiments and measurements by Grabke and
    Rieke~\cite{gr}, the vertical width of the band reflecting the
    reported error bars. The remaining lines are data from
    electrochemical permeation experiments, assessed by Kiuchi and
    McLellan~\cite{km} and measured by Nagano~{\it et~al.};~\cite{nh}
    and measurements using gas and electrochemical permeation by
    Hayashi~{\it et~al.}~\cite{hh2} The extent of each data set
    represents the temperature range over which the assessment or
    measurements are reported. The triangles show theoretical results
    from centroid molecular dynamics calculations using a classical
    interatomic potential and are taken from ref.~\cite{k1}.}
\begin{center}
\includegraphics[scale=0.4,viewport=26 0 719 720,clip]{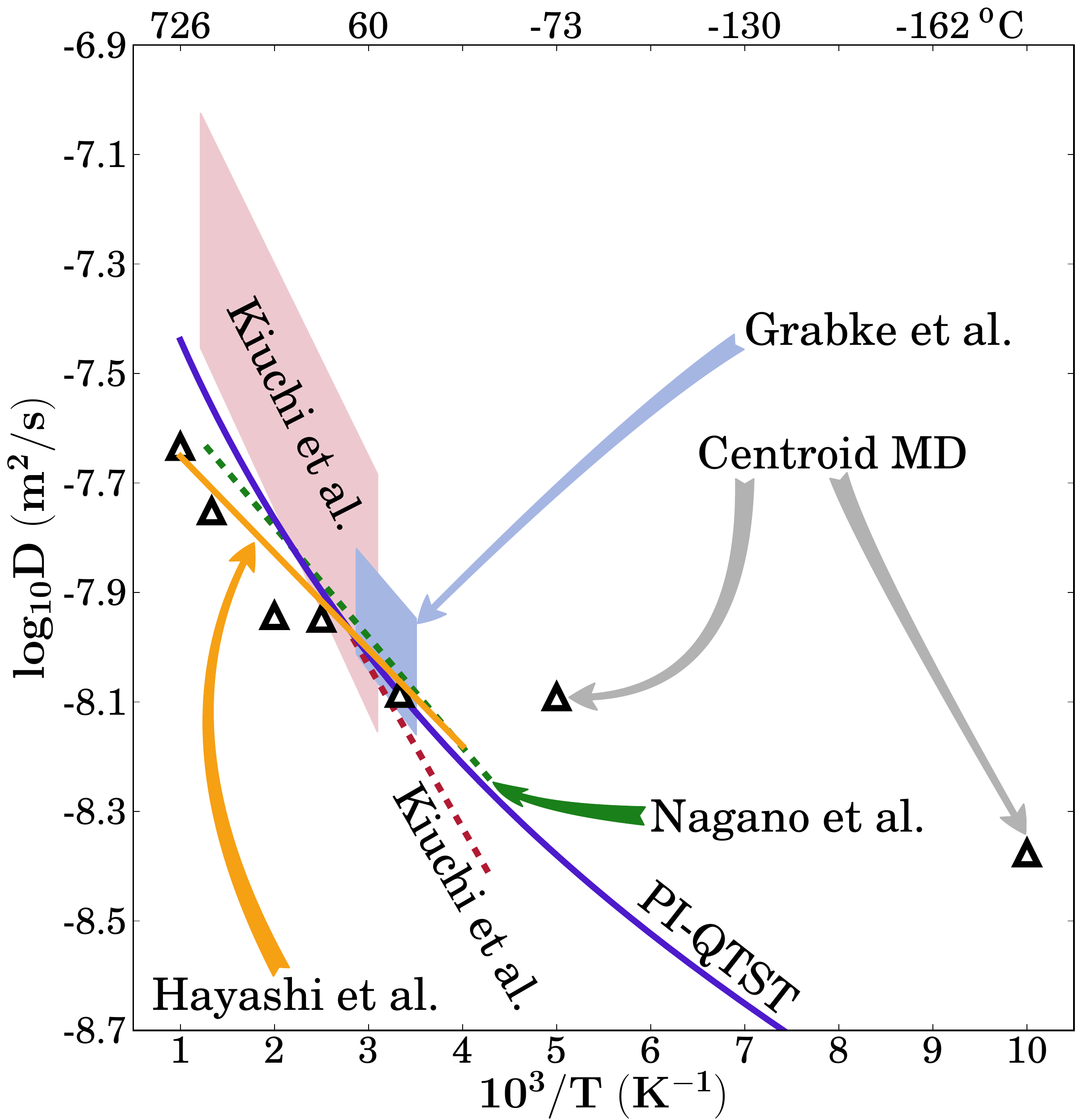}
\end{center}
\end{figure*}

Exactly the same formula applies in the quantum transition state
theory if one simply replaces the classical partition functions
with their quantum mechanical counterparts~\cite{Voth93}. The way
to calculate quantum mechanical partition functions is to use
Feynman's path integral formalism~\cite{Feynman72}. A very
readable account can be found in ref~\cite{GillanNato}. Feynman
shows that the partition function at temperature $T$ of a single
particle moving in a potential $V(x)$, in one dimension for
simplicity, is
\begin{equation}
Z = \int{\cal D}x(\tau)\,e^{S/\hbar}
\label{PathIntegral}
\end{equation}
Here $\tau$ has units of time, although it is a measure of the
(inverse) temperature. The use of the symbol ${\cal D}x(\tau)$ means
that the integral is taken over all ``paths'' starting at $\tau=0$ and
ending at $\tau=\beta\hbar$~\cite{Feynman72}. In general a path is
such that the particle starts out at position $x_1$, say, at $\tau=0$,
that is, having infinite temperature, and travels to position $x_2$ at
the temperature $T$ at which we require the partition
function. However for the purpose of calculating $Z$ the only paths
included in the integral are those for which the position of the
particle is the same at $\tau=0$ and $\tau=\beta\hbar$ so that these
are {\it closed} paths in that sense. Therefore
equation~(\ref{PathIntegral}) is the integral of $\exp(S/\hbar)$
taken over all positions $x$ and all paths for which the particle
travels from $x$ and back again in an ``imaginary'' time $\beta\hbar$
which is determined by the temperature of interest. In
equation~(\ref{PathIntegral})
$$
S[x(\tau)] = \int_0^{\beta\hbar}{\rm d}\tau
    \left[-\frac{1}{2}m\dot x^2(\tau) - V(x(\tau))\right]
$$
has the form of a classical action, except for the sign of the
kinetic energy, which here is a second derivative with respect to
{\it imaginary} time, $\tau=\beta\hbar$~\cite{Gillan87JPC}.

If one discretises the closed paths into $P$ segments, then the
partition function can be approximated to~\cite{GillanNato}
\begin{equation*}
 Z \approx \left(\frac{mP}{2 \pi\beta\hbar^2}\right)^\frac{P}{2} 
\int dx_1...dx_P \,\,\exp \left\{ -\beta\sum_{s=1}^P \left[
    \frac{1}{2}\frac{mP}{\beta^2\hbar^2}
(x_{s+1}-x_s)^2+P^{-1} V(x_s) \right] \right\}
\label{PI}
\end{equation*} 
where again this is an integral over all closed paths, which means in
the discretised case that $x_{i+P}=x_{i}, \hskip 3pt\forall
i$~\cite{Cao94}. What is astonishing about this formula is that
the partition function of a {\it quantum mechanical} particle is
equivalent to the {\it classical} partition function belonging to a
necklace of $P$ beads connected by harmonic springs of stiffness
$mP/\beta^2\hbar^2$. Each bead feels a potential energy $V(x)/P$. The
numerical estimate converges to equation~(\ref{PathIntegral}) when the
discretisation parameter $P$ is chosen to be large enough.

Armed with the approximate partition functions, these can be
substituted for the classical functions in
equation~(\ref{rate_constant}) to obtain the rate constant,
$\kappa^{\rm QTST}$, in the quantum transition state
theory~\cite{Voth93}. Finally the diffusivity can be obtained from the
Einstein formula, assuming an uncorrelated random walk among the
tetrahedral sites of the bcc lattice~\cite{ChristianI},
$$
D = \frac{1}{6} z  R^2 \kappa^{\rm QTST}
$$
where, again, $R^2=a^2/8$ and $z=4$ is the number of neighbouring
tetrahedral sites. In figure~\ref{figure2} is shown the result of the
calculation compared to experiment~\cite{Katzarov13}. There are two
points to note. One is that the path integral quantum transition state
theory (PI-QTST) predicts non Arrhenius behaviour with an up turning
of the diffusivity at low temperature. This is consistent with the
point made earlier that at very low temperature the proton tunnels
freely between potential wells; in the limit of zero Kelvin the
particle is delocalised. A second point is that the quantum mechanical
treatment is essential: the classical theory predicts an activation
energy that is about one third too large at room
temperature~\cite{Katzarov13}.

Even though it is necessary to treat the motion of the protons in
steel quantum mechanically this does not mean that the classical
minimum energy path is unimportant. On the contrary, as has been
indicated above, the quantum mechanical approach to tunnelling
requires knowledge of the classical activation energy. Moreover it
must be understood that the computation of the {\it classical\/}
minimum energy path (that is, treating the nuclei as classical) is
still a quantum mechanical calculation if tight binding or density
functional theory is used to calculate the interatomic forces.

\begin{figure*}
  \caption{\label{fig_vacMEP} Calculated minimum energy paths for
    hydrogen migration to a vacancy in \aFe. Each curve, labelled $n$H
    corresponds to a case where already $n-1$ H~atoms have been
    absorbed at the vacancy. The curve labelled ``bulk'' is the same
    minimum energy path as shown in figure~\ref{figure1}. The
    ``reaction coordinate'' along the abscissa shows the fractional
    length of the total path, or length in configuration space of the
    ``nudged elastic band''~\cite{Henkelman00}, which is different for
    each case. For example the two end points (zero and one on the
    abscissa) illustrated in terms of their atomic structure for the
    case of five H~atoms are drawn in figure~\ref{fig_vacn5}. }
\begin{center}
\includegraphics[scale=0.5,viewport=56 179 545 636,clip]{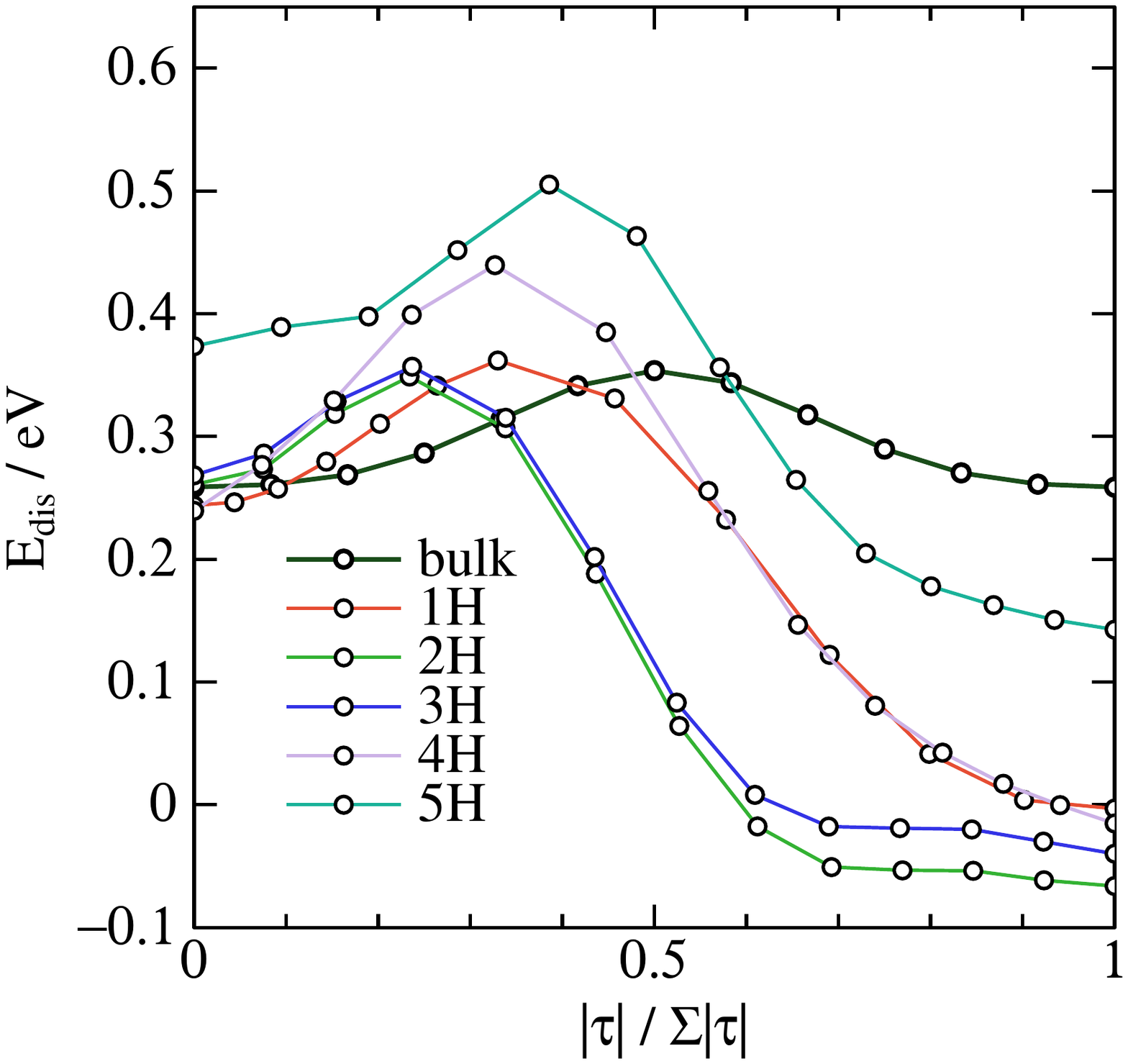}
\end{center}
\end{figure*}

\subsubsection{Hydrogen trapping at vacancies}

\begin{figure*}
  \caption{\label{fig_vacn5} Atomic structures of the two end points
    of the minimum energy path labelled ``5H'' in
    figure~\ref{fig_vacMEP}. At the start of the path, there are four
    H~atoms already trapped by the vacancy occupying near-tetrahedral
    sites at the faces of the cube, and there is one proton in a
    near-bulk tetrahedral site in the unit cell to the left. At the
    end of the path (the lower figure) that proton has migrated into a
    fifth near-tetrahedral site at the vacancy that has now trapped
    five protons.}
\begin{center}
\includegraphics[scale=0.3,viewport=56 152 407 850,clip]{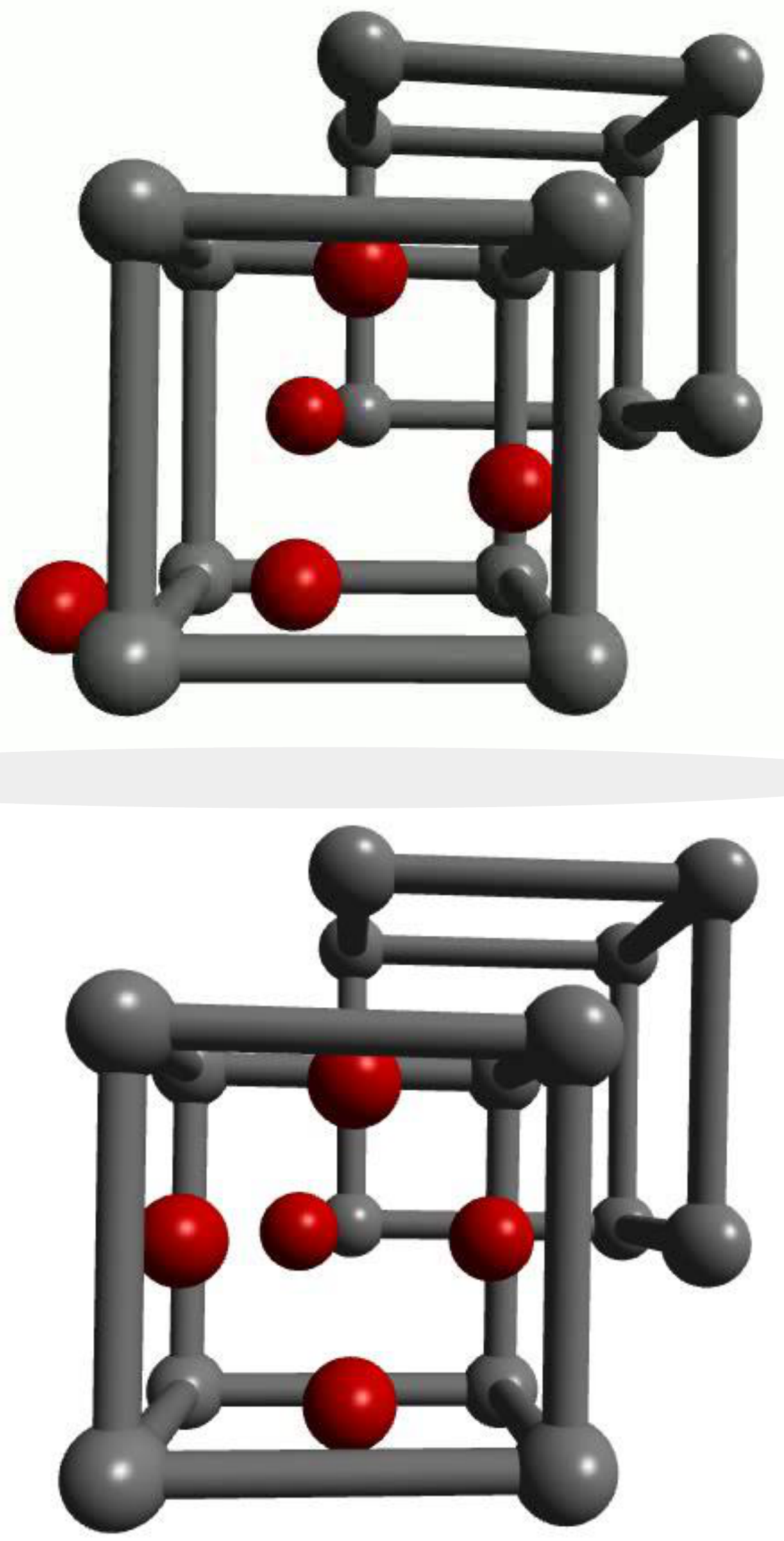}
\end{center}
\end{figure*}

We can extend the study of hydrogen in the perfect bcc lattice to
consider its interaction with vacancies. This is of great
technological importance as hydrogen induced superabundance of
vacancies is thought to be a mechanism of hydrogen
embrittlement~\cite{Nagumo03,Nagumo04}. Figure~\ref{fig_vacMEP} shows
minimum energy paths for H~atoms migrating from a neighbouring unit
cell into a unit cell containing an iron vacancy in \aFe. We show
successive paths as first one, then two, then three, up to the case
where five H~atoms have been absorbed by the vacancy. For
illustration, the atomic structures of the initial and final state are
shown for the case ``5H'' in figure~\ref{fig_vacn5}.

The energy plotted in the ordinate of figure~\ref{fig_vacMEP} is the
``dissolution energy'', namely the total energy of the \aFe\ crystal
containing hydrogen compared to an equivalent perfect crystal and the
same amount of hydrogen in the H$_2$ gaseous
state~\cite{EAC09,Paxton10}. In the case of hydrogen migration in
perfect bulk \aFe, labelled ``bulk'' in figure~\ref{fig_vacMEP}, this
is always positive, reflecting the fact that hydrogen is very
insoluble in \aFe. Figure~\ref{fig_vacMEP} shows that hydrogen atoms
can lower their potential energy by becoming absorbed at a vacancy and
in the case where two or three H~atoms are trapped the dissolution
energy is negative. This observation is extremely significant in
supporting the notion that vacancies act as traps for
H~atoms. Hydrogen traps deeper than about 0.5~eV are often said to be
irreversible, that is one would not expect any significant thermal
desorption at temperatures up to 300${^\circ}$C~\cite{Szost13}. Under
this classification we see that the vacancy in ferrite acts as a {\it
  reversible} trap for hydrogen. Figure~\ref{fig_vacMEP} also shows
that up to five H~atoms can be transferred exothermically from bulk
tetrahedral sites to a single vacancy. This result is consistent with
experiment~\cite{Iwamoto99} and with DFT
calculations~\cite{Tateyama03}. On the other hand a quantitative
comparison is less convincing. Myers~{\it et al.}~\cite{Myers89} have
assigned trap depths of 0.43~eV and 0.63~eV to respectively 3--6 and
1--2 deuterium atoms trapped at a vacancy in \aFe. In that case the
trend is the same as in the theory, but the magnitudes are at odds
with it. In particular, we find in figure~\ref{fig_vacMEP} that the
trap depth is about the same, 0.25--0.3~eV, for all cases up to four
H~atoms. The so called superabundance of vacancies~\cite{Iwamoto99} is
an example of what has been called the ``defactant'' effect by
Kirchheim~\cite{Kirchheim07a}. This is in analogy with the usage {\it
  surfactant} to describe a chemical which acts to lower the free
energy of a surface; here the hydrogen species acts to lower the free
energy of formation of the vacancy defect by becoming trapped within
it.

\subsection{Carbon in \aFe}

One might well ask whether there is a defactant effect at work due to
interactions between carbon atoms and vacancies in steels and whether
this has implications for heat treatment and microstructures in
martensite and bainite. As yet this question is
unresolved~\cite{Francisca}, although there have been some very
interesting DFT calculations by F\"orst {\it et al.}~\cite{Forst06}
which can answer this question. These authors showed that a number of
defect complexes exist in steels comprising carbon interstitials and
vacancies. According to their calculations, the two predominant point
defects at 160${^\circ}$C, apart from alloying impurities, are the
carbon interstitial in octahedral sites and a vacancy {\it having two
  carbon atoms bound to it}. This rather startling result plainly has
huge implications for the understanding of both carbon and self
diffusivity in \aFe.

Recently a tight binding model for carbon in magnetic \aFe\ has been
demonstrated and this can be used to good effect to reproduce this
result and to give an interpretation~\cite{Paxton13}. 

\begin{table}
\newdimen\digitwidth
\setbox0=\hbox{\rm0}
\digitwidth=\wd0
\vskip -12pt
\caption{\label{table} Properties of interstitial carbon in iron
  calculated with a tight binding model~\cite{Paxton13} and compared
  to density functional
  calculations~\cite{Domain01,Jiang03,Forst06,Lau07,Kabir10} and
  experiment~\cite{Seeger98}. The properties are, 
  in eV, the vacancy formation ($H_{\rm Vac.}^{\rm F}$) and
  migration ($H_{\rm Vac.}^{\rm M}$) energies in pure \aFe, the
  carbon migration energy in \aFe, $H^{\rm M_{\alpha}}_{\rm C}$, and
  carbon migration energies in $\gamma$-Fe. These last two correspond respectively to a single hop
  between octahedral sites,  $H^{\rm M_{\gamma}}_{\rm C}$(d), squeezing the carbon atom in between two
  nearest neighbour Fe atoms and a double hop via an intermediate
  tetrahedral site,  $H^{\rm M_{\gamma}}_{\rm C}$(tet). The final two
  columns show the binding energies of one and two carbon atoms to a
  vacancy in \aFe\ (see the text for details).}
\centerline{ \vbox{ \catcode`~=\active
    \def~{\kern\digitwidth}
    \def\tablerule{\noalign{\smallskip\hrule\smallskip}}
    \def\doubletablerule{\noalign{\smallskip\hrule\vskip
        1pt\hrule\smallskip}} \hrule\vskip 1pt\hrule
\medskip
\halign{
  \hfil#\hfil &\quad  \hfil#\hfil &\quad
  \hfil#\hfil &\quad  \hfil#\hfil &\quad
  \hfil#\hfil &\quad  \hfil#\hfil &\quad
  \hfil#\hfil &\quad  \hfil#\hfil \cr
& $H_{\rm Vac.}^{\rm F}$ & $H_{\rm Vac.}^{\rm M}$ &
 $H^{\rm M_{\alpha}}_{\rm C}$ & 
 $H^{\rm M_{\gamma}}_{\rm C}$(d) &
 $H^{\rm M_{\gamma}}_{\rm C}$(tet) &
 $E_B$(1) & $E_B$(2)  \cr
\tablerule
TB &     1.6        &  0.8       &  0.87   & 1.00  & 1.48  & 0.47  & 1.50  \cr
DFT &    2.0        & 0.65--0.75 &  0.81   & 0.63  & 2.11  & 0.64  & 1.65  \cr
expt. &  1.61--1.75 & 1.12--1.34 &         &       &       &       &       \cr
}
\medskip
\hrule\vskip 1pt\hrule
}
}
\end{table}

\begin{figure*}
  \caption{\label{fig_aCdiff} Atomic structure snapshots of the
    calculated minimum energy path for carbon diffusion in perfect
    \aFe. It can be seen that in the first image the elastic dipole
    points roughly left to right as the crystal is locally
    tetragonally distorted by the carbon atom occupying the irregular
    octahedron of the perfect bcc lattice. As the point defect moves
    the dipole rotates into an up and down direction in the figure;
    that is, the dipole originally along the $x$-axis rotates
    90${^\circ}$ into the $z$-axis. The elastic interaction between
    point defects will be such as to cause these to align in analogy
    with the alignment of spins in a ferromagnetic
    material~\cite{Zener48,Nowick63}. It is this effect that leads to
    the tetragonality of martensite and bainite~\cite{Leslie,Bhadeshia}.}
\begin{center}
\includegraphics[scale=1,viewport=0 600 335 765,clip]{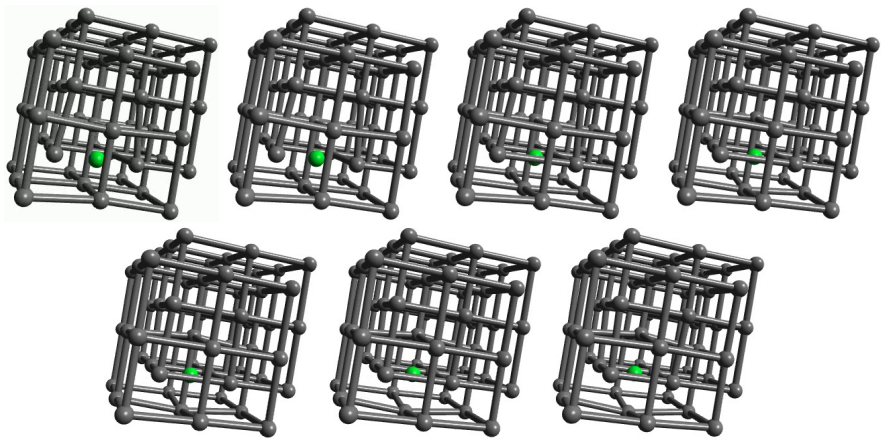}
\end{center}
\end{figure*}

\begin{figure*}
  \caption{\label{fig_gCdiff} Atomic structure snapshots of the
    calculated minimum energy path for carbon diffusion in
    $\gamma$-Fe. Surprisingly the C~atom does not take an intermediate hop
  via a neighbouring tetrahedral site for which the activation energy
  is 2.1~eV (table~\ref{table}); instead it forces its way
  through the nearest neighbour Fe--Fe bond at the so called high
  energy ``d''-saddle point~\cite{Paxton13} for which the energy
  barrier is very much smaller---0.63~eV.}
\begin{center}
\includegraphics[scale=1,viewport=28 634 320 767,clip]{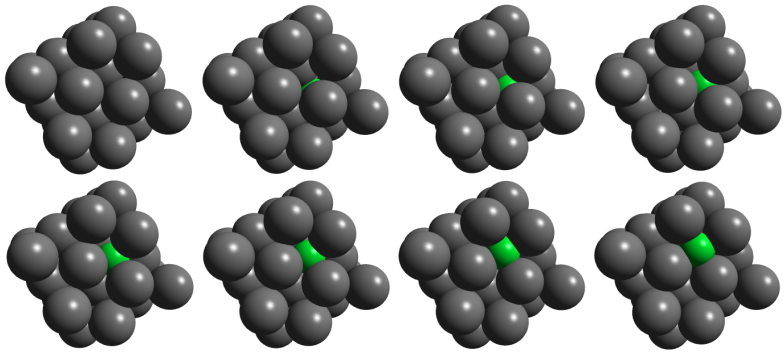}
\end{center}
\end{figure*}

Some calculated data relevant to carbon in steel are shown in
table~\ref{table} in which tight binding results are compared to DFT
and to experiment. The TB model employs the Slater-Stoner theory of
collinear itinerant magnetism~\cite{Paxton08} and so non collinear
effects are not captured; this may lead to small errors in the
treatment of $\gamma$-Fe~\cite{Abrikosov99}. The experimental vacancy
formation and migration energies are, surprisingly, still very
controversial; here we follow the discussion by Seeger and quote his
results~\cite{Seeger98}. The TB migration energies for C in ferrite
and austenite are in quite good agreement with DFT results. C
migration in \aFe\ is illustrated in figure~\ref{fig_aCdiff}. This
shows rather clearly the local tetragonal strain intoduced into the
lattice by the C~interstitial. This can be modelled as a double
Boussinesq force without couples~\cite{Eshelby51} and gives rise to an
elastic dipole which can interact with a second rank tensor, stress
field, in close analogy with a electric dipole which responds to a
vector, electric field~\cite{Nowick63}. Figure~\ref{fig_aCdiff} shows
rather clearly how the elastic dipole reorients through 90${^\circ}$
after each hop. In austentite the carbon atom has a choice of
migrating between octahedral sites via an intermediate tetrahedral
site or it may take the direct route by forcing its way through the
centre of the nearest neighbour Fe--Fe bond. As the energies in
table~\ref{table} show the carbon, rather surprisingly, chooses the
second option; the minimum energy path is illustrated in
figure~\ref{fig_gCdiff}.

Finally, we can turn to the question of the binding of carbon
interstitials to a vacancy in \aFe. The data in the last two columns
of table~\ref{table} show a quantity $E_B$ defined by Becquart~{\it et
  al.}~\cite{Becquart07}. This is the total energy of a defect complex
compared to the energy of the constituent point defects when widely
separated. So, for example, $E_B$(2) is the energy of an isolated
vacancy and two carbon interstitals {\it minus} the energy of two
carbon atoms bound to a vacancy. A positive value means that the
complex is more stable, in this case by 1.65~eV, than its separated
components. In other words the crystal gains 1.65~eV (159~kJ/mol) when
two carbon atoms become bound to a bare vacancy. We have seen in the
case of hydrogen that firstly the H~atom does not occupy the actual
vacant site---that entails a large energy
penalty~\cite{Tateyama03}. Instead the H~atoms occupy near-tetrahedral
sites in the cube faces surrounding the vacancy. Secondly we have seen
that a vacancy in \aFe\ will absorb up to five H~atoms out of solution
exothermically; moreover the configuration with two or three trapped
H~atoms actually has a negative heat of solution with respect to
hydrogen gas. Carbon also does not occupy the actual vacant site,
substitutional phases of carbon in Fe have very large, positive heats
of formation~\cite{Paxton13}. Instead carbon atoms {\it could} occupy
up to the six octahedral positions at the centres of the faces of the
cube surrounding the vacant site~\cite{Forst06}. Of the six possible
Fe-vacancy--C-interstitial complexes, the cases of two and three carbon
atoms have the greatest binding energy~\cite{Forst06} as indicated in
table~\ref{tableC}, hence the prediction that the vacancy plus two
carbon atoms is the next most predominiant defect after the isolated
carbon interstitital in Fe--C alloys~\cite{Forst06}.

\begin{table}
\newdimen\digitwidth
\setbox0=\hbox{\rm0}
\digitwidth=\wd0
\vskip -12pt
\caption{\label{tableC} Binding energies, $E_B$($n$), in eV as defined by
  Becquart~{\it et al.}~\cite{Becquart07} for up to six carbon atoms
  to a vacancy in \aFe\ as calculated using DFT by F\"orst~{\it et
    al.}~\cite{Forst06}.
 These atoms occupy the octahedral sites at the
centres of the six faces bounding the vacant site which thereby form the
points of a regular octahedron (see figure~1 in
ref.~\cite{Forst06}). The first two columns are the same as already
reported in table~\ref{table}. Also shown, for comparison, are the
equivalent data for hydrogen both using density functional theory~\cite{Tateyama03} and
tight binding~\cite{Paxton10}. Note that a vacancy will trap more
hydrogen than carbon atoms: six as opposed to four, the seventh would occupy the
vacancy itself and this is not a trap site. The TB is in reasonable
agreement with the DFT, but it wrongly predicts that the vacancy {\it
is} a trap site. The H~atoms do not occupy the octahedral sites as do
C; for the details of the atomic structure, see ref~\cite{Paxton10}.}
\medskip
\centerline{ \vbox{ \catcode`~=\active
    \def~{\kern\digitwidth}
    \def\tablerule{\noalign{\smallskip\hrule\smallskip}}
    \def\doubletablerule{\noalign{\smallskip\hrule\vskip
        1pt\hrule\smallskip}} \hrule\vskip 1pt\hrule
    \medskip
    \halign{
      \hfil# &\quad  \hfil#\hfil &\quad  \hfil#\hfil &\quad
      \hfil#\hfil &\quad  \hfil#\hfil &\quad  \hfil#\hfil &\quad  \hfil#\hfil &\quad
      \hfil#\hfil &\quad  \hfil#\hfil \cr
      $n$ & = & 1 & 2 & 3 & 4 & 5 & 6 & 7 \cr
      \tablerule
      C (DFT)&  & 0.64 & 1.65 & 1.78 & 1.31 & --1.66 & --7.98  & \cr
      H (DFT)&  & 0.56 & 1.17 & 1.57 & 1.85 &  ~2.18 &  ~2.16  & --0.52 \cr
      H (TB) &  & 0.32 & 0.65 & 0.91 & 1.07 &  ~1.22 &  ~1.19  & ~0.71 \cr
    }
    \medskip
    \hrule\vskip 1pt\hrule
  }
}
\end{table}

\begin{figure*}
  \caption{\label{fig_dimer} Four possible atomic
    structures~\cite{Domain04,Forst06} of the carbon dimer bound to a
    vacancy in \aFe\ calculated by static relaxation using tight
    binding interatomic forces~\cite{Paxton13}. All four are local
    minima in the potential energy. Note that in all but case~(a) the
    two C~atoms have joined together to form a C$_2$ ``dimer
    molecule'' inside the vacancy. The global minimum is predicted to
    be the structure labelled~(d) because that offers the preferred
    four-fold coordination of the carbon atom, exactly as it
    experiences in the ethane molecule, although here of course the
    bonds are to Fe, not H~atoms. See the text for a discussion.}
\begin{center}
\includegraphics[scale=0.8,viewport=30 22 304 297,clip]{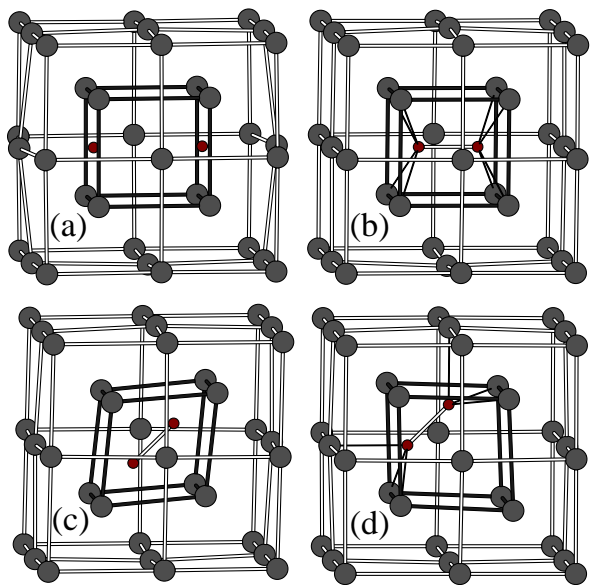}
\end{center}
\end{figure*}

The reason for the large stability of the vacancy plus two carbon
atoms can be seen in figure~\ref{fig_dimer} which shows four possible
configurations of the two carbon atoms bound to the vacancy that have
been proposed~\cite{Domain04,Forst06}. The structure labelled~(a) is
perhaps the most obvious, but in fact it has the highest potential
energy. What is predicted to happen is the that the carbon atoms
approach each other to form a covalent bond, whose bond length
is~1.43~\AA\ which is the same as the bond distance in diamond. The
structure is therefore that of a C$_2$ dimer bound to the vacancy. The
structures labelled~(b)--(d) in figure~\ref{fig_dimer} correspond to
possible orientations of the dimer and in fact, as predicted both in
DFT and in tight binding it is the structure labelled~(d) that has the
lowest potential energy~\cite{Paxton13}. The reason for this can be
seen in figure~\ref{fig_dimer}. In this orientation each carbon atoms
makes four bonds, one to the other carbon atom to form the dimer and
three further bonds of equal length, 3.65~\AA, to neighbouring
Fe~atoms. In this way the carbon atom satisfies its requirement of
four fold coordination. This is an example in which a quantum
mechanical theory is expected to be essential if the phenomenon is to
be correctly described, because the bonding is achieved by the well
known $sp$-hybridisation of the carbon atomic orbitals~\cite{Coulson}.

\section{Concluding remarks}

The aim of this paper has been to show how electronic structure
calculations can be made that have direct relevance to physical
metallurgy. The state of the art is the density functional theory, but
it has been demonstrated that an abstraction into a tight binding
model can be both reliable and predictive. This means that in the
future it will not always be necessary to rely on the somewhat
uncontrollable approximations attendant upon simulations using
classical potentials. In particular it is now possible to make
atomistic models of total energy and interatomic force in magnetic
iron and steel. In the present case this has been applied to the
calculations of atomic and electronic structures of hydrogen and
carbon interstitials in iron. This has resulted in new insights into
the nature of vacancies as traps for hydrogen and carbon and it is
expected that this will have future impact in the understanding of
hydrogen embrittlement and processes in the heat treatment of steel,
leading ultimately to the design of new materials.

\section*{Acknowledgements}

This work was supported under the programme MultiHy (Multiscale
Modelling of Hydrogen Embrittlement in Crystalline Materials, Grant
Number 263335, {\tt www.multihy.eu}) by the European Union's 7th
Framework Program under the theme “Nanosciences, Nanotechnologies,
Materials and new Production Technologies”.

\def\JPCM{J.~Phys.:~Condens.~Matter}
\def\PRB{Phys.~Rev.~B}
\def\PRL{Phys.~Rev.~Lett.}

\bibliographystyle{mst}


\end{document}